\documentclass[11pt]{article}
\usepackage[dvips]{graphicx}
\usepackage{mathrsfs}
\usepackage{amsthm} 
\usepackage{amsmath}
\usepackage{amssymb}
\usepackage{amsfonts}
\usepackage{float}
\usepackage{yhmath}
\usepackage{wrapfig}
\usepackage[dvips]{color}
\usepackage[dvips]{colortbl}
\usepackage{cite}
\usepackage{array}
\usepackage{tabularx}
\usepackage[sectionbib]{chapterbib}
\usepackage{threeparttable}
\usepackage{chemarrow}
\usepackage{framed}
\usepackage{ascmac}
\definecolor{shadecolor}{gray}{0.90}
\usepackage[small, bf, labelsep=quad]{caption}
\usepackage{cases}

\newcommand{\BL}[1]{\textcolor{#1}{$-\hspace{-0.6mm}-$}}

\newcommand{\I}{I}
\newcommand{\II}{I\hspace{-0.5mm}I}

\definecolor{mycolor1}{rgb}{1,1,0.7}
\definecolor{mycolor2}{rgb}{0.9,1,1}
\definecolor{mycolor3}{cmyk}{0,0,0,0.113}
\definecolor{mycolor4}{cmyk}{0.086,0,0,0}

\begin{document}

\begin{center}
\Huge{Size Exponents of Branched Polymers}
\end{center}
\vspace*{-3mm}
\begin{center}
\LARGE{Extension of the Isaacson-Lubensky Formula and the Application to Lattice Trees}
\end{center}

\vspace*{10mm}
\begin{center}
\large{Kazumi Suematsu\footnote{\, The author takes full responsibility for this article.}, Haruo Ogura$^{2}$, Seiichi Inayama$^{3}$, and Toshihiko Okamoto$^{4}$} \vspace*{2mm}\\
\normalsize{\setlength{\baselineskip}{12pt} 
$^{1}$ Institute of Mathematical Science\\
Ohkadai 2-31-9, Yokkaichi, Mie 512-1216, JAPAN\\
E-Mail: suematsu@m3.cty-net.ne.jp, ksuematsu@icloud.com  Tel/Fax: +81 (0) 593 26 8052}\\[3mm]
$^{2}$ Kitasato University,\,\, $^{3}$ Keio University,\,\, $^{4}$ Tokyo University\\[15mm]
\end{center}

\hrule
\vspace{3mm}
\noindent
\textbf{\large Abstract}: 
Branched polymers can be classified into two categories that obey the different formulae:
\begin{equation}
\nu=
\begin{cases}
\hspace{1mm}\displaystyle\frac{2(1+\nu_{0})}{d+2} & \hspace{3mm}\mbox{for polymers with}\hspace{2mm}\displaystyle\nu_{0}\ge\frac{1}{d+1}\hspace{10mm}\text{(\I)}\\[3mm]
\hspace{5mm}2\nu_{0}& \hspace{3mm}\mbox{for polymers with}\hspace{2mm}\displaystyle\nu_{0}\le\frac{1}{d+1}\hspace{10mm}\text{(\II)}
\end{cases}\notag\label{2022A-11}
\end{equation}
for the dilution limit in good solvents. The category {\II} covers the exceptional polymers having fully expanded configurations. On the basis of these equalities, we discuss the size exponents of the nested architectures and the lattice trees. In particular, we compare our preceding result, $\nu_{d=2}=1/2$, for the $z$=2 polymer having $\nu_{0}=1/4$ with the numerical result, $\nu_{d=2}\doteq 0.64115$, for the lattice trees generated on the 2-dimensional lattice. Our conjecture is that while both the conclusions in polymer physics and condensed matter physics are correct, the discrepancy arises from the fact that the lattice trees are constructed from less branched architectures than the branched polymers having $\nu_{0} = 1/4$ in polymer physics. The present analysis suggests that the 2-dimensional lattice trees are the mixture of isomers having the mean ideal size exponent of $\bar{\nu}_{0}\doteq0.32$.
\vspace{0mm}
\begin{flushleft}
\textbf{\textbf{Key Words}}: Branched Polymers/ Lattice Trees/ Ideal Size Exponent/ Size Exponent/
\normalsize{}\\[3mm]
\end{flushleft}
\hrule
\vspace{10mm}
\setlength{\baselineskip}{14pt}

\section{Introduction}
The Isihara formula is a mathematical expression that measures the spatial distance from the center of gravity to the $p$th monomer:
\begin{equation}
\vec{r}_{Gp}=\vec{r}_{1p}-\frac{1}{N}\sum_{p=1}^{N}\vec{r}_{1p}\label{Isihara-1}
\end{equation}
and is a basis to study the configurational statistics of macromolecules. In spite of the apparent mathematical simplicity, the Isihara formula has a rich physical extension. The most prominent property of this formula is that the vector, $\vec{r}_{Gp}$, can be decomposed into the grand sum of all bond vectors, $\{\vec{l}_{q}\}$, that constitute the polymer, namely,
\begin{equation}
\vec{r}_{Gp}=\frac{1}{N}\sum_{q=2}^{N} c_{q}(p)\,\vec{l}_{q}\label{Isihara-2}
\end{equation}
Starting from $q=2$ is necessary since an $N$ polymer has $N-1$ bonds. Let all bonds have an equal length, $|\vec{l}|=l$. The mean square of the radius of gyration is written, quite generally, in the form:
\begin{equation}
\left\langle s_{N}^{2}\right\rangle=\frac{1}{N}\sum_{p=1}^{N}\left\langle\vec{r}_{Gp}\cdot\vec{r}_{Gp}\right\rangle=\frac{1}{N}\sum_{p=1}^{N}\left[\sum_{q=2}^{N}\frac{c_{q}(p)^{2}}{N^{2}}+\sum_{i\neq j}^{N} \frac{c_{i}(p)c_{j}(p)}{N^{2}}\left\langle\vec{e}_{i}\cdot\vec{e}_{j}\right\rangle\right]l^{2}\label{Isihara-3}
\end{equation}
where $\vec{e}_{i}$ denotes the unit vector of bond $i$\cite{Isihara, Weiss, Redner, Kazumi2}. Eq. (\ref{Isihara-3}) is independent of architectures and, therefore, holds whether a polymer is linear or branched.

In the preceding works\cite{Kazumi2}, we put forth that, for $g\rightarrow\infty$, the unperturbed sizes of branched polymers vary in proportion to the generation number, $g$, of the main backbone. We may thus write the above expression in the scaling form:
\begin{equation}
\langle s_{g}^{2}\rangle\propto g^{2\lambda}\,l^{2}\hspace{5mm}(g\rightarrow\infty)\label{2022A-4}
\end{equation}
Eq. (\ref{2022A-4}) can be justified by the fact that all polymers have the same mathematical expression [Eq. (\ref{Isihara-3})] for $\left\langle s_{N}^{2}\right\rangle$, independently of the architectures, and the scaling law is well-established for linear polymers\cite{Flory, Isaacson}.

The relationship between $g$ and $N$ can be calculated by making use of the structural information about individual polymers. For the nested structures introduced in the preceding works, this relationship is given by the recursion relation that has the solution, $g\doteq N^{\frac{1}{z}}$ for a large $g$, which leads to
\begin{equation}
\langle s_{N}^{2}\rangle\propto N^{2\frac{\lambda}{z}}\,l^{2}\hspace{5mm} (z=1, 2, 3, \cdots)\label{2022A-5}
\end{equation}
Hence we have the relation among the exponents:
\begin{equation}
\nu=\frac{\lambda}{z}\label{2022A-6}
\end{equation}
By Eq. (\ref{2022A-4}), it is obvious that $\lambda=1/2$ for the freely jointed polymers, and we have 
\begin{equation}
\nu_{0}=\frac{1}{2z}\label{2022A-7}
\end{equation}
Combining Eq. (\ref{2022A-6}) with Eq. (\ref{2022A-7}), we have further
\begin{equation}
\nu=2\lambda\nu_{0}\label{2022A-8}
\end{equation}
Now the restraining condition inherent to the nesting structures has been removed altogether, and we have gained the general relationship (\ref{2022A-8}) among the exponents, $\nu$, $\lambda$, and $\nu_{0}$.

\begin{figure}[h]
\begin{center}
 \includegraphics[width=9cm]{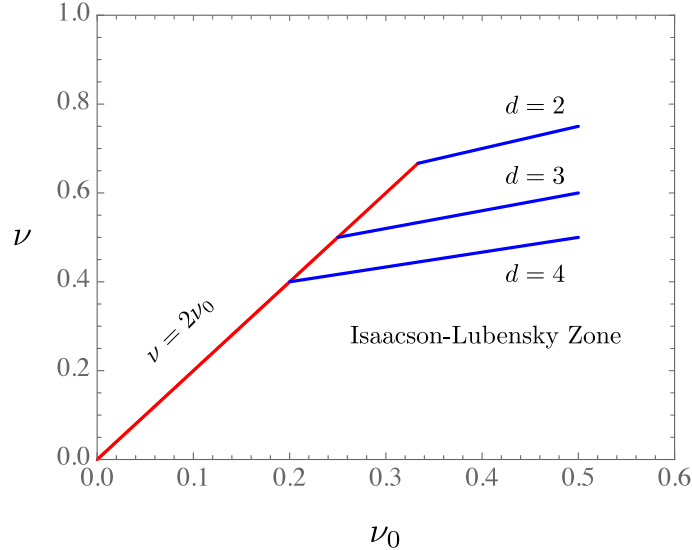}
 \caption{$\nu_{0}$ vs. $\nu$ diagram for an isolated polymer in the good solvent. Red solid-line (\BL{red}) is by the equation $\nu=2\nu_{0}$; blue-solid lines (\BL{blue}) is the Isaacson-Lubensky prediction.}\label{ExponentFormulae}
 \end{center}
\vspace{-3mm}
\end{figure}

Clearly, the radius of gyration can not grow beyond the own size of a molecule, and one must have $\lambda\le 1$. Substituting this into Eq. (\ref{2022A-8}), we have the inequality:
\begin{equation}
\nu\le 2\nu_{0}\label{2022A-9}
\end{equation}
Using the inequality (\ref{2022A-9}), we can examine the applicability range of the Isaacson-Lubensky formula, $\nu=2(1+\nu_{0})/(d+2)$, the formula for isolated polymers in good solvents; the result is
\begin{equation}
\nu_{0}\ge\frac{1}{d+1}\label{2022A-10}
\end{equation}
According to the discussion through Eq. (\ref{2022A-8}) to Eq. (\ref{2022A-10}), polymers that violate Eq. (\ref{2022A-10}) must have $\lambda=1$\cite{MILNER, Grest}, so those polymers must satisfy $\nu=2\nu_{0}$. It is apparent, from the viewpoint of the exponents, that the isolated polymers in good solvents may be divided into two categories that obey the different formulae:
\begin{equation}
\nu=
\begin{cases}
\hspace{1mm}\displaystyle\frac{2(1+\nu_{0})}{d+2} & \hspace{3mm}\mbox{for polymers with}\hspace{2mm}\displaystyle\nu_{0}\ge\frac{1}{d+1}\hspace{10mm}\text{(\I)}\\[3mm]
\hspace{5mm}2\nu_{0}& \hspace{3mm}\mbox{for polymers with}\hspace{2mm}\displaystyle\nu_{0}\le\frac{1}{d+1}\hspace{10mm}\text{(\II)}
\end{cases}\label{2022A-11}
\end{equation}
Eq. (\ref{2022A-11}) is plotted in Fig. \ref{ExponentFormulae}. Within our knowledge, the known largest $\nu_{0}$ is $1/2$ of linear molecules. $\nu$ goes down, starting from $\nu_{0}=1/2$, linearly along the Isaacson-Lubensky line (\BL{blue}), terminates precisely at $\nu_{0}=\frac{1}{d+1}$, and then, again, goes down along the $\nu=2\nu_{0}$ line (\BL{red}).

\begin{figure}[h]
\vspace{0mm}
\begin{center}
 \includegraphics[width=15.5cm]{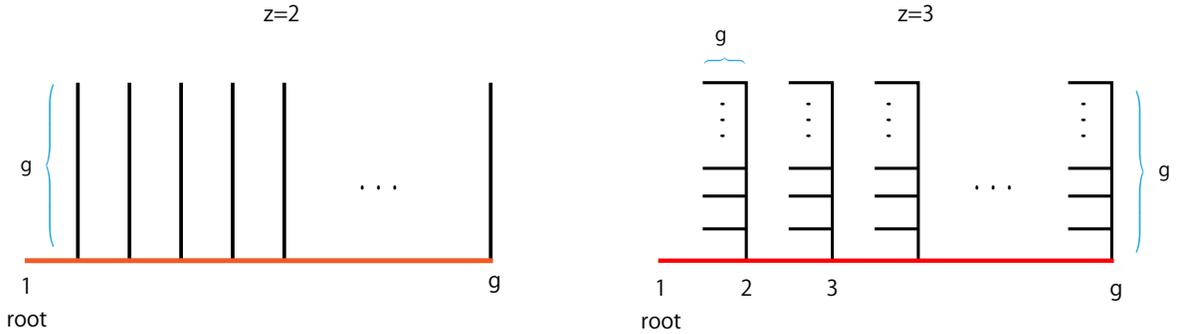}
 \caption{Nested branched architectures. $z$ denotes the depth of the nest. The $z$=2 polymer has $\nu_{0}=1/4$, and the $z$=3 polymer has $\nu_{0}=1/6$\cite{Kazumi2}.}\label{z2z3polymers}
 \end{center}
 \vspace{0mm}
\end{figure}

\begin{shaded}
\vspace{-5mm}
\subsubsection*{\textsf{[}Examples\textsf{]}} 
The $z$=3 polymer (Fig. \ref{z2z3polymers}) introduced in the preceding paper\cite{Kazumi2} has $\nu_{0}=1/6$, so it belongs to the second category {\II} in Eq. (\ref{2022A-11}) for $d=3$. We have thus $\nu=2\cdot\frac{1}{6}=1/3$, in agreement with the direct algebraic calculation of $\langle s_{N}^{2}\rangle$ on the simple cubic lattice (this polymer can not be embedded in the two-dimensional space!). On the other hand, the $z$=2 polymer has $\nu_{0}=1/4$, so it is on the boundary between the categories {\I} and {\II} for $d=3$. Thus, both the formulae ({\I} and {\II}) yield the same result $\nu=1/2$, the well-established value in polymer physics\cite{Isaacson, Parisi, Daoud, Seitz, Hsu, Kazumi2, Rensburg} (Figs. \ref{ExponentFormulae} and \ref{z2z3polymers}).
\end{shaded}

\section{On the Ideal Exponent of Lattice Trees}
As has been well known, the lattice trees are not identical, in the composition and the isomer ratio, with the randomly branched polymers synthesized in the laboratory\cite{Seitz, Kazumi1}. This is because of the realization of the steric hindrance in the lattice models and the resulting reduction in functionality, $f$. Nevertheless, the lattice models are very useful, since they provide accurate numerical values to check the prediction of polymer physics. The most prominent benefit might be the information on the size exponent, $\nu$.

Let us focus our attention on the size exponents on the square lattice ($d=2$). The determination of the exponent is a long-standing issue in condensed matter physics and has not yet been solved rigorously. It was found that the values ($\nu_{d=2}\approx 0.64$) observed so far for the lattice trees and animals\cite{Family, Seitz, Derrida, Hsu, Rensburg} conflict with the value ($\nu_{d=2}=1/2$) rigorously proven in polymer physics for the branched polymers having $\nu_{0}=1/4$\cite{Kazumi2, Everaers}. For the lattice trees, a very reliable estimate of $\nu_{d=2}=0.64115$ was put forth by Jensen\cite{Jensen} some time ago. Now that such convincing data have been provided, we are ready to address this open question.

First, we must accept the following facts: (i) both polymer physics and lattice theories (simulations, series expansion, and renormalization calculations) deal with the branched polymers; the former handles the randomly branched polymer or the branched polymer with fixed configuration (say, the $z$=2 polymer), and the latter the mixture of various isomers which are assumed to have attained the asymptotic limit with respect to the exponents; (ii) both the estimates are correct, or almost correct (the former has been rigorously solved for the $z$=2 polymer on the square lattice, and the latter's numerical estimates, all, fall on near $0.64$\cite{Family, Seitz, Derrida, Jensen, Hsu, Rensburg}). If these are the case, we must conclude that the two approaches deal with different branched molecules.

Then, our question is, what kinds of samples have we so far examined? One useful way to answer this question will be to identify the ideal exponent, $\nu_{0}$, of the polymers handled so far. In polymer physics, the samples were always the randomly branched polymer or the $z$=2 polymer, the ideal exponents of which have been rigorously proven to be $\nu_{0}=1/4$\cite{Zimm, Dobson, Kazumi2}. On the other hand, in the lattice theories, while $\nu$'s for the self-avoiding polymers have been computed, with great precision, devising ingenious technical methods, none of the papers, strangely enough, have confirmed the ideal exponents, $\nu_{0}$, for the lattice trees\cite{Jensen, Hsu, Rensburg} in question.

\subsection{Evaluation of the Ideal Exponent}
To date, the Isaacson-Lubensky formula was the only one mathematical expression that predicts the size exponent algebraically, and it served as the standard for checking the validity of the calculation results. To this formula, the new equation of the category {\II} was added for the exceptional polymers that satisfy $\nu_{0}\le\frac{1}{d+1}$. Note that Eq. (\ref{2022A-11}) is a formula that predicts $\nu$ from $\nu_{0}$, but at the same time, it is a formula that predicts $\nu_{0}$ from $\nu$.

Let us apply the most reliable value, $\nu_{d=2}\doteq 0.64115$, estimated for the self-avoiding lattice trees to Eq. (\ref{2022A-11}), and we have
\begin{description}
\item[1.] with the Isaacson-Lubensky formula
\begin{equation}
\nu_{0}=0.2823\label{2022A-12}
\end{equation}
which contradicts the constraint condition: $\nu_{0}\ge 1/3\, (=0.33\dots)$, so this estimation must be renounced.
\item[2.] with $\nu=2\nu_{0}$
\begin{equation}
\nu_{0}=0.320575\label{2022A-13}
\end{equation}
which is compatible with the constraint condition $\nu_{0}\le 1/3$.
\end{description}
Within the framework of Eq. (\ref{2022A-11}), it is suggested that the lattice trees generated on the square lattice, as a mixture of various isomers, should have the mean ideal exponent, $\bar{\nu}_{0}\doteq 0.32$, slightly short of the critical value 1/3. In contrast to the value, 1/4, as has been assumed so far, the lattice trees should be constructed from less branched molecules at $d=2$. The result is in harmony with our conclusion in the preceding paper, which showed that the deepest degree of branching at $d=2$ is, at most, that of the $z$=2 polymer with $\nu_{0}=1/4$. It follows that the branched trees on the square lattice should consist of the mixture of polymers having $\nu_{0}=1/2$ to 1/4, so that those should have the mean ideal exponent that falls on $1/4\le\overline{\nu}_{0}\le1/2$, consistent with the evaluation of Eq. (\ref{2022A-13}).

The numerical value deduced in Eq. (\ref{2022A-13}), at present, remains a conjecture. However, if every isomer's structure is identified, the verification might be possible, for instance, by means of the Kramers theorem\cite{Kramers, Kazumi2}:
\begin{equation}
\langle s_{N}^{2}\rangle_{0}=\frac{l^{2}}{N^{2}}(N-1)\sum_{k=1}^{N-1}\omega_{k}\,k(N-k)\label{2022A-14}
\end{equation}
where $\omega_{k}$ is a weight for the $(k, N-k)$ pair.

Finally, it might be important to emphasize again that both the theories in polymer physics and condensed matter physics are mathematically correct. Only, those theories have dealt with different systems in different fields, independently of each other.  In retrospect, the fortuitous coincidence between the Isaacson-Lubensky prediction, $5/8=0.625$, and the estimates, $\approx 0.64$, by the lattice theories seems to have so long clouded our thinking on the issue of the exponent, $\nu_{d=2}$. As discussed in the preceding papers, this value is precisely 1/2 for the branched polymers with $\nu_{0}=1/4$.

\vspace{5mm}

\end{document}